\documentclass[a4paper,fleqn]{cas-dc}
\renewcommand{\printorcid}{}

\usepackage[authoryear]{natbib}
\usepackage{arydshln}
\usepackage{amsmath}
\usepackage{subcaption}
\usepackage{caption}

\def\tsc#1{\csdef{#1}{\textsc{\lowercase{#1}}\xspace}}
\tsc{WGM}
\tsc{QE}
\tsc{EP}
\tsc{PMS}
\tsc{BEC}
\tsc{DE}

\begin{document}
\let\WriteBookmarks\relax
\def\floatpagepagefraction{1}
\def\textpagefraction{.001}
\shorttitle{End-to-end acoustic modelling for phone recognition of young readers}
\shortauthors{Lucile Gelin et~al.}


\title [mode = title]{End-to-end acoustic modelling for phone recognition of young readers}                   



\author[1,2]{Lucile Gelin}[prefix=]

\author[2]{Morgane Daniel}[]

\author[1]{Julien Pinquier}[%
   prefix=
   ]
\author[1]{Thomas Pellegrini}[%
   prefix=
]
\address[1]{IRIT, Paul Sabatier University, CNRS, Toulouse, France}
\address[2]{Lalilo, France}

\begin{abstract}
Automatic recognition systems for child speech are lagging behind those dedicated to adult speech in the race of performance.
This phenomenon is due to the high acoustic and linguistic variability present in child speech caused by their body development, as well as the lack of available child speech data. Young readers speech additionally displays peculiarities, such as slow reading rate and presence of reading mistakes, that hardens the task.
This work attempts to tackle the main challenges in phone acoustic modelling for young child speech with limited data, and improve understanding of strengths and weaknesses of a wide selection of model architectures in this domain. 
We find that transfer learning techniques are highly efficient on end-to-end architectures for adult-to-child adaptation with a small amount of child speech data. 
Through transfer learning, a Transformer model complemented with a Connectionist Temporal Classification (CTC) objective function, reaches a phone error rate of 28.1\%, outperforming a state-of-the-art DNN-HMM model by 6.6\% relative, as well as other end-to-end architectures by more than 8.5\% relative. 
An analysis of the models' performance on two specific reading tasks (isolated words and sentences) is provided, showing the influence of the utterance length on attention-based and CTC-based models. 
The Transformer+CTC model displays an ability to better detect reading mistakes made by children, that can be attributed to the CTC objective function effectively constraining the attention mechanisms to be monotonic.  
\end{abstract}




\maketitle

\section{Introduction}



\sloppy
Speech recognition systems, enabled by the advent of high computational capacities, are nowadays widely used in various applications: human-machine interaction, home assistantship, education, tutoring, entertainment. 
Being more and more exposed to numerical technologies, children are becoming an important part of the public using these systems. However, although the speech recognition accuracy is approaching perfection on adult speech, the performance on child speech is far and away from it~\citep{Potamianos2003-RRO, Shivakumar20-TLF}. 
Tackling the challenges brought by child speech for speech recognition thus becomes essential to level up global accuracy levels.

Numerous research projects have been conducted on the differences between adult and child speech~\citep{Potamianos2007-ARA, Mugitani12-DOV,Lee99-AOC,Gerosa2006-AAA}. They have shown important acoustic, linguistic and prosodic variability in child speech. 
Acoustic variability is caused by developmental changes in the early years of life, in particular in the speech production apparatus. 
The small size of the vocal tract induces shifted fundamental and formant frequencies in high frequencies, slowly transferring to lower frequencies during growth, and reaching adult level around 15 years old~\citep{Mugitani12-DOV}. 
This evolution causes high inter-speaker variability across age groups. Research shows high intra-speaker acoustic variability in the spectral domain, particularly for young children, due to the stabilisation of pitch control occurring only around age 8~\citep{Lee99-AOC}. 
Linguistic and pronunciation variabilities can be attributed to the slow development of articulatory mechanisms, which are not fully developed at ages 5-7. 
Bad positioning of the tongue and lips can also cause phonological errors among young children, which show to disappear with age~\citep{Fringi15-EOP}.
Finally, prosodic variability is caused by developing and limited knowledge of the language: prosodic events, such as lower duration of phones, false starts, hesitations and breathing have been found more frequently from 8 to 10-year-old children's speech in comparison with older age groups~\citep{Potamianos1998-SDS}.

A common application of speech recognition for children has educational purposes: reading tutors. Learning to read is indeed a challenge for young children, who might need extra help to master it. 
Several projects have been implemented over the years~\citep{Mostow2001-ETT,Bolanos11-FOR, Proenca2018-AAO,Godde2017-EOR}, applied to different languages, age groups and reading tasks. 
Lalilo\footnote{\url{https://www.lalilo.com/}} provides an online reading assistant for 5-8 year old children, featuring reading aloud exercises where children record themselves reading and get feedback on their reading. 
It aims at enabling extra practice in reading aloud, which is often not allowed a lot of time in class because of practical issues, while providing feedback and support in areas of difficulty for the student. 
Speech recognition for children learning to read is an arduous task: non-proficient readers' speech contains many disfluencies and reading mistakes that can be laborious to detect automatically. 
The usage of reading assistants in classroom environments adds up difficulties due to the common presence of babble noise on the recordings, sometimes concealing the target student's speech.

\section{Related works}

Prior studies on ASR for child speech have shown that the performance is below that of latest systems dedicated to adult speech~\citep{Potamianos2003-RRO, Shivakumar20-TLF}. 
ASR for child speech is a relatively recent research domain, that has expanded with the creation of dedicated events, for instance the Workshop on Child Computer Interaction (WOCCI)\footnote{\url{https://sites.google.com/view/wocci}} and the Child Speech Recognition Challenge organised during the IEEE Spoken Language Technology Workshop~\citep{Yu2020-SLT}.
A study on grade-specific ASR performance shows significant gaps of word error rate (WER): 26.9\%, 14.6\%, 10.5\% and 5.1\% among children of age 5, 6, 7 and 8+, confirming that the younger the children, the lower the accuracy~\citep{Yeung18-OTD}. 
The same study shows promising results by training children-specific and even grade-specific ASR models, as long as sufficient data are available to learn despite the important acoustic variability in child speech.

Due to limited available children data in most languages, deep neural network-based systems only recently started to be exploited. \citet{Serizel14-DNN} train a hybrid Deep Neural Network - Hidden Markov Model (DNN-HMM) system on adult and children mixed data, that serves as initialisation to train a children-adapted model with only child speech. 
An amount of seven hours of child speech training data, ages 7-13, enables them to reach a 12.4\% phone error rate.
For a children language learner application, \citet{Metallinou14-UDN} present a DNN-HMM that, even trained on less data, surpasses Gaussian Mixture Model - Hidden Markov Model (GMM-HMM) systems. 
Their best-performing model reaches a 19.3\% WER with 430 hours of child speech training data.
\citet{Shivakumar20-TLF} provide valuable insights on acoustic modelling for child speech recognition with DNN-HMM, by training models on adult speech, then adapting them with a dataset of 91 hours of 6-14 year old children's speech, reaching a 17.8\% WER. They also carry a deep analysis of the influence of several parameters such as the age of the speakers and the quantity of child data. 

End-to-end architectures have shown to equal (or outperform) hybrid architectures on a broad range of ASR datasets \citep{Karita2019-ACS}.
Their major advantage is to discard the need of pre-segmenting training data and post-processing outputs, directly predicting  the final label sequence, which considerably simplifies the process of decoding utterances. 
This unification of the whole training process in a single step, prevents from potential behavioural mismatches between modules that have been trained separately. In the same objective of simplicity, end-to-end systems mostly act at the character level, since it eliminates the process of converting the word-level transcriptions into phones, thus the need of a pronunciation dictionary.
A first step towards end-to-end ASR was made by \citet{Graves2006-CTC} with the Connectionist Temporal Classification (CTC) objective function, used with success with a model based on recurrent neural networks (RNN) on the TIMIT phone recognition task.
Sequence-to-sequence (seq2seq) architectures were then presented, with encoder-decoder structures \citep{Sutskever2014-STS,Chorowski14-E2E,Chorowski2015-ABM,Lu2015-ASO,Chiu2018-SSR}. 
The Listen, Attend and Spell (LAS) architecture~\citep{Chan2015-LAS} put forward innovative methods, using an attention mechanism to link the encoder and decoder modules, based on RNNs, and bypass the conditional independence assumption made by the CTC.
Transformer architectures~\citep{Vaswani2017-AIA, Dong2018-STR} made the most of attention mechanisms by replacing the RNNs usually contained in both encoder and decoder by multi-head self-attention modules, enabling better versatility and faster computing. 
All seq2seq architectures (LAS, Transformer...) can also be trained with a combination of cross-entropy (CE) and CTC objective functions, as presented in \citet{Watanabe2017-HCA} and \citet{Karita2019-ITB}. This method aims at constraining the attention to be more monotonic, as well as helping the system to converge faster. 
A recent study demonstrated the superior ASR performance of the Transformer trained with CE+CTC objectives on diverse individual corpora and on multilingual tasks, in comparison with a LAS-like and hybrid DNN-HMM models~\citep{Karita2019-ACS}.

Due to the limited available child speech data, very few studies present child speech recognition with end-to-end models. Methods though exist, like Transfer Learning (TL) that consists in training an acoustic model on a large out-of-domain dataset, then retraining it with a small in-domain corpus to retain part of the acquired knowledge and adapt it to the application domain. 
It showed significant performance improvement with hybrid DNN-HMM approaches for adult-to-children transferring~\citep{Shivakumar20-TLF,Tong2017-TLF, Qian2016-IDN} applications. TL was also used on end-to-end architectures for low resource languages \citep{Cho2018-MSS, Tong2017-MTC}.

As for end-to-end models for child ASR, \citet{Andrew2015-AMW} show improvement on child speech with a CTC-based system jointly trained on very large quantities of mixed adult and child speech data. 
Usage of seq2seq models for child speech recognition is a new research subject, as show the extremely recent communication of technical reports on this matter~\citep{Ng2020-CUH,Chen2020-DAF}. Both studies use the same dataset, containing 59 hours of Mandarin speech from 7 to 11-year-old children.
The former uses transfer learning for adult-to-child adaptation, and reports a 23.6\% character error rate (CER) with a CE+CTC multi-objective RNN-based seq2seq model. On the other hand, the latter mixes adult and child speech, and announces 18.8\% CER with a convolutional-augmented Transformer model and data augmentation techniques.
An even more recent study reports a comparison of diverse end-to-end systems for automatic child speech recognition, but they use a substantial amount of data (223 hours) from children that are quite older (9-11 years old) than in our study, and focus on conversational speech over read speech \citep{Shivakumar2021-E2E}. They obtain a 9.2\% CER and 16.0\% WER on matched test data with a self-attention-based encoder trained with a CTC objective function. On test data with a wider age range (6-16), they obtain a 33.6\% WER with a Time-Depth Separable Convolution Network, with significant performance degradation for 5-8 years old children due to acoustic differences across ages and higher intra-speaker variability in young children's speech.
While these recent works do character recognition, as end-to-end systems were designed for, in this work we stick to modelling phones instead, in agreement with our final application: pronunciation assessment of children learning to read.
Furthermore, these studies lack a detailed evaluation of strengths and weaknesses of seq2seq models in comparison with state-of-the-art hybrid DNN-HMM approaches, when applied to child speech in a low-resource scenario.

Our work attempts to furnish valuable insights in this new research domain. We apply a wide catalogue of end-to-end ASR systems to child speech, to:
\begin{itemize}
    \item Compare recent end-to-end models (RNN, LAS, Transformer) to a baseline hybrid DNN-HMM model for phone recognition;
    \item Show the efficiency of transfer learning on end-to-end models for adult-to-child adaptation, even with very low amounts of child data;
    \item Analyse the influence of different components of end-to-end architectures, such as the attention mechanism or the connectionist temporal classification (CTC) objective function, on the performance;
    \item Evaluate the models on different reading tasks: isolated words and sentences;
    \item Assess the impact of CE+CTC multi-objective training for evaluation of utterances containing reading mistakes;
\end{itemize}

This article is organised as follows. Section~\ref{sec:corpus} presents the adult and child speech material used in this study, and offers a detailed description of typical mistakes that children learning to read do.
Section~\ref{sec:TL_description} provides comprehensive explanation of our transfer learning strategy.
Section~\ref{sec:acoustic_models} presents the acoustic models that we apply to child speech in this paper: the baseline, a hybrid DNN-HMM model, as well as several end-to-end models.
Section~\ref{sec:results} shows the validation of our models on adult speech, and studies the results obtained on child speech. 
Finally, section~\ref{sec:discussion} offers analyses that provide further insights on end-to-end acoustic models' behaviour when confronted with different reading tasks and reading levels.

\section{Speech material}
\label{sec:corpus}
We use two datasets of French speech: the \textit{Common Voice} adult corpus\footnote{Corpus available at: \url{https://voice.mozilla.org/fr}}, and an in-house child speech corpus, hereinafter called \textit{Lalilo}. Tables~\ref{tab:info_corpus_CV} and~\ref{tab:info_corpus_Lalilo} display general information on these data.


\subsection{Adult dataset: Common Voice}
The Common Voice corpus is created through a participatory online platform, where everyone can record himself reading sentences. 
Being composed of recordings with different equipment in different environments, it is thus particularly adapted to our reading speech task in classrooms. Nonetheless, since speakers usually record themselves alone in their computer room, the data does not contain babble noise and has a high mean signal-to-noise ratio (SNR).
In French, the Train, Valid and Test sets we used for these experiments contain respectively 148.9, 2.4 and 7.2 hours of speech (see Table~\ref{tab:info_corpus_CV}). 
The Test set has been designed to maximise the number of distinct speakers, while keeping the same average time per speaker as in the Valid set. Each recording is validated by two annotators, thereby the corpus contains a limited quantity of miscues.

\begin{table}[!htb]
    \caption{Information on Common Voice (CV) adult speech dataset}
    \label{tab:info_corpus_CV}
    \begin{tabular}{l | c c c}
        \toprule
        \textbf{Dataset} & Train & Valid & Test \\
        \midrule
        \textbf{Duration} (h) & 148.9 & 2.4 & 7.2 \\
        \textbf{Speakers} & 1276 & 372 & 1113 \\
        \textbf{Avg time} & \multirow{2}{*}{420.1} & \multirow{2}{*}{23.5} & \multirow{2}{*}{23.4} \\
        \textbf{per spk} (s) & & & \\ \hdashline[1pt/1pt]
        \textbf{SNR} (dB) & & & \\
        \hspace{0.1cm} Mean & 34.4 & 34.3 & 34.3 \\
        \hspace{0.1cm} Std & 14.7 & 14.5 & 14.7 \\
        \bottomrule
    \end{tabular}
\end{table}

\subsection{Child dataset: Lalilo}
The Lalilo corpus contains recordings of Kindergarten to second grade children, aged 5-8, reading aloud isolated words and sentences. These two tasks are commonly given by teachers to beginning readers, according to their reading level. Detailed information is displayed in Table~\ref{tab:info_corpus_Lalilo}. 
The recordings have been gathered either in person directly in schools, or through a reading aloud exercise in the Lalilo web platform. In the first case, the environmental conditions are somewhat clean: a good-quality microphone is used, and the noise level is controlled. 
In the second case, however, teachers usually let a small group of students play on the platform, with reduced supervision, which inevitably implies a highly variable level of babble noise on the recordings. Because schools do not always benefit from headsets, computer built-in microphones are sometimes used, which record a wide range of ambient noise.
Table~\ref{tab:info_corpus_Lalilo} displays the mean SNR for each train and test sets, and shows that the Lalilo corpus contains more noise than the Common Voice corpus (see Table ~\ref{tab:info_corpus_CV}).

The training set contains sentences and isolated words, with an equal proportion of isolated words and words from sentences. Similarly, we use a validation set that contains both isolated words and sentences. 
The reading assistant offers children to read either isolated words or sentences, according to their reading level, thus we constructed a test set that can be split in two, each corresponding to one task. They are named Test W (for words) and Test S (for sentences) in Table~\ref{tab:info_corpus_Lalilo}.

Phonetically transcribing with exactitude what has been read by non-reader children is a challenging task, due to articulation mechanisms in the process of being acquired, potential hesitations and stammering, or even the presence of acoustically non-existing phones. 
Although it is not an impossible task, it is very time-consuming and costly, and we have chosen not to invest in phone-level transcriptions for the whole corpus at the moment. Accordingly, only correctly pronounced and fluently read utterances that are identical to the prompted text, were included in the training and validation data for both the acoustic and language models. 
Only the test data, containing words and sentences with reading mistakes, has been phonetically transcribed to enable the computation of performance scores. The transcriptions have been made manually by two human judges, and recordings have been discarded in case of disagreement.

\begin{table}[!htb]
    \caption{Information on Lalilo child speech dataset. The "Test W" and "Test S" respectively designate the isolated words and sentences test sets. The "\% correct read" designates the percentage of phones belonging to an utterance that contains no reading mistake.}
    \label{tab:info_corpus_Lalilo}
    \begin{tabular}{l | c c c | c c}
        \toprule
        \textbf{Dataset} & Train  & Valid & Test & Test W & Test S \\
        \midrule
        \textbf{Duration} (h) & 13.0 & 0.41 & 1.32 & 0.84 & 0.48 \\
        \textbf{Speakers} & 3014 & 459 & 685 & 425 & 262 \\
        \textbf{Avg time} & \multirow{2}{*}{15.2} & \multirow{2}{*}{3.2} & \multirow{2}{*}{6.9} & \multirow{2}{*}{7.0} & \multirow{2}{*}{6.7} \\
        \textbf{per spk} (s) & & & & \\ \hdashline[1pt/1pt]
        \hdashline[1pt/1pt]
        \textbf{\% correct read} & 100 & 100 & 57.6 & 48.5 & 64.0\\
        \hdashline[1pt/1pt]
        \textbf{SNR} (dB) & & & & & \\
        \hspace{0.1cm} Mean & 20.9 & 20.1 & 20.6 & 22.2 & 22.8 \\
        \hspace{0.1cm} Std & 13.1 & 12.7 & 12.6 & 13.2 & 12.1 \\
        \bottomrule
    \end{tabular}
\end{table}

\subsection{Reading learners mistakes}
\label{sec:typical_mistakes}
Mistakes done by reading learning children are diverse and sometimes unique, which makes the manual annotation of data quite difficult for human experts, and can cause all the more confusion for an ASR system. 
Reading mistakes can be categorised in two main categories, word-level and sentence-level, then into several subcategories. Understanding the different types of reading mistakes can help analyse the system's behaviour when encountering these.

Reading mistakes at the word level can usually enter in one (or both) of the following categories:
\begin{itemize}
    \item \textbf{Hesitation}: presence of intra-word silence(s) due to the child hesitating when reading;
    \item \textbf{Substitution}: a phone is substituted by another. \\Example: \textit{cat} is read \textit{cut};
    \item \textbf{Insertion}: a phone is inserted inside a word.\\Example: \textit{cat} is read \textit{cart};
    \item \textbf{Deletion}: a phone is deleted from a word. \\Example: \textit{cat} is read \textit{at}.
\end{itemize}

For substitutions, insertions and deletions, the resulting word can either exist or not.

Reading mistakes can also occur at the sentence level, like repetitions of on or several words, false starts (the child attempts to read the word but the word is not read entirely), or hesitations between words.
Word-level mistakes are more difficult to detect (for both humans and machines) than sentence-level mistakes, as they usually concern one or two phones: in particular, substitutions and insertions of phones are easy to miss. 
Furthermore, the data we gathered comes from the Lalilo reading assistant that adapts the difficulty of exercises to each child's reading level: children who were asked to read isolated words were more susceptible to having a lower global reading level than children who had to read sentences, causing the isolated word recordings to contain very laborious readings. 
The Lalilo Test W set thus offers a particularly arduous task.


A word is considered as erroneously read when it contains at least one of the reading mistakes listed above. 
A sentence is considered as such when it contains at least one erroneous word, or a sentence-level reading mistake. Table~\ref{tab:info_corpus_Lalilo} displays the percentage of phones included in correctly read utterances in Lalilo Test W and S sets: respectively 48.5\% and 64.0\% of phones belong to utterances that contain no reading mistake.
These percentages are in agreement with the recordings gathered on the Lalilo platform: children reading at lower levels are given words to read, thus more than half isolated words contain a reading mistake, while more advanced students are giving sentences to read and statistically make fewer reading mistakes.

\section{Tackling the low-resource configuration}
\label{sec:TL_description}

Transfer learning (TL) can be done on many ASR applications, such as adaptation between languages \citep{Abad2020-CLT,Tong2017-MTC,Cho2018-MSS}, between different speech types, such as broadcast news and conversational speech in \citet{Abad2020-CLT}, or between native and non-native speech \citep{Duan2020-CLT}. According to the application, the available quantity of target data, or the similarity between the source and target domains, it will be applied differently.
In our case, adult-to-child adaptation, the source model is a model trained on adult speech, while the adaptation data is child speech data. This particular adaptation is very sensitive to the amount of target data and the target children's age, since their vocal apparatus and speech quality are so different than adults, and vary greatly during children's growth~\citep{Shivakumar20-TLF}.
We follow this article's findings: overall, they find that the best adaptation configuration is either to adapt the whole network or to adapt two layers at the bottom and two layers at the top, both strategies obtaining the exact same score. 
However, for young (6-8 years old) children, they show that the high complexity and variability in acoustic and prosodic characteristics of their speech necessitate more parameters (hence layers) to be fully captured, suggesting that the first option should be chosen. 
A minimum amount of adaptation data is nevertheless required (\textasciitilde10 hours): below, adapting the whole network would hurt performance due to noise introduced by high variability. 
When sufficient adaptation data are provided, the network is able to learn young child speech variability. In these conditions, adapting the whole model performs better than only some layers. 
Our training child corpus containing approximately 13 hours of speech from very young children (5-8 years old), we follow the advice and choose to apply transfer learning on the whole source model in our experiments.

For each acoustic model architecture presented in the next section, we trained a model on adult speech to serve as a source model, then used TL to adapt this model with our child speech data.

\section{Phone-level acoustic models}
\label{sec:acoustic_models}
This study aims at comparing the performance of state-of-the-art end-to-end and baseline hybrid DNN-HMM ASR systems when applied to the challenging task that is child speech recognition at the phone level, with a low quantity of data and a noisy environment.
This section will present the different acoustic models that we review in this work. We will train, with each architecture, two separate models: 
\begin{itemize}
    \item Adult model, trained on \textit{Common Voice} data with almost 150 hours;
    \item TL model, using the adult model and adapting it with \textit{Lalilo} child speech data (\textasciitilde13 hours).
\end{itemize}

\subsection{TDNNF-HMM: the baseline}

Acoustic models for speech recognition were originally built from a Hidden Markov Model (HMM), linked to either a single Gaussian distribution, or a more refined Gaussian Mixture Model (GMM). 
The latter models the state-output distribution in an unsymmetrical and multi-modal manner, in opposition to the former, enabling the accurate modelling of speaker, accent or gender differences in speech~\citep{Gales2008-AHM}. 
With the advent of deep neural networks (DNN), the GMM-HMM models were progressively replaced by hybrid DNN-HMM acoustic model architectures. 
This change of paradigm introduces a slight change in probability computing: while GMM directly provide the posterior probability $P(x_t|s)$ of an observation $x$ at time $t$ given a state $s$, used by the HMM, the DNN computes $P(s|x_t)$, from which $P(x_t|s)$ can be computed with Bayes' rule. 
Although GMM-HMM models are not anymore used for acoustic modelling, they keep the purpose of generating the alignments on which DNN-HMM models are trained.

Time-Delay Neural Networks (TDNN), introduced for phone recognition by \citet{Waibel89-PRU}, shown to be particularly adapted for ASR, with their ability to represent relationships between acoustic events in time while providing time-invariance of the features learnt by the network. 
It introduces delays in the weighted sum of inputs which, in practice, are implemented as spatially expanded units over a certain number of frames. 
Context width varies depending on the layers: bottom layers learn short duration acoustic-phonetic characteristics, while top layers learn more complex features of longer duration. For instance, TDNNs have been used with success for vowel recognition on children's speech in a small-resource language~\citep{Yong11-SIV}.

We use as a baseline a Factorised Time-Delay Neural Network (TDNNF)~\citep{Povey18-SOL}, a refinement of TDNN models, with a Lattice-Free Maximum Mutual Information (LF-MMI) criterion~\citep{Povey16-PST}. These models are currently the state-of-the-art in speech recognition, and have shown to reach an 11.7\% WER on 55 hours of child speech data~\citep{Wu2019-AIA}.


We use the Kaldi toolkit~\citep{Povey11_TKS} to train our TDNNF acoustic models, which follow available recipes\footnote{\url{https://frama.link/script-TDNNF}}$^,$\footnote{\url{https://frama.link/script-TDNNF-TL}}. 
A 3-state GMM-HMM monophone model provides alignments to train the TDNNF: it gave better results than using triphone GMM models, due to a small number of occurrences of each triphone in the child speech training dataset, as well as the presence of non-existing phones (reading mistakes) in the testing set. The network has a chain model architecture~\citep{Povey16-PST}, and is trained with the LF-MMI criterion, similarly to~\citet{Vesel2013-SDT}. 
The network contains 12 TDNNF layers of dimension 1024 as hidden layers, each comprised of one TDNN layer of dimension 1024 and two bottleneck layers of dimension 128. 
The input features are 40-dimensional Mel-frequency cepstral coefficients (MFCC) with Cepstral Mean and Variance Normalisation (CMVN), as do~\citet{Bayerl2019-ACH}, as well as classical Kaldi recipes. Adult and TL models are trained on 990 and 89 epochs respectively, both with a learning rate of 5e-4 and a $l_2$ regularisation rate of 1e-2. The adult and TL models have a size of 7.6M parameters, and respectively took 46 and 2.1 hours to train on a single GTX 2080 Ti GPU.

\subsection{End-to-end systems}

Hybrid approaches have made possible the use of recurrent neural networks --which are highly efficient for modelling time series~\citep{Graves2013-SPW}-- for sequence labelling, by combining them with HMMs that provide pre-segmentation of the input sequence for training and post-transformation into the output label sequence. 
However, the training process of DNN-HMM models is quite complex, with the need of generating alignments of each speech utterance and then using lattices to infer a label sequence. Furthermore, its functioning relies on several parts (acoustic model, language model, pronunciation model), which are trained independently with different objectives, causing behavioural mismatches between components.

End-to-end architectures aim at simplifying and unifying the training process of a speech recognition system, by encompassing all components into a single neural network. 
In this objective, end-to-end models are usually trained on characters, which can be directly derived from words, instead of phones that can only be obtained with a pronunciation dictionary. In this work, we will stick with phones, since it suits better for modelling the grapheme-phoneme correspondence that children learn when starting reading.  

Given a sequence of input speech features $X=(x_{1},...,x_{T})$, and the corresponding phone-label output sequence $Y = (y_{1},...,y_{N})$, with $T$ the number of frames in the speech utterance and $N$ the length of the phone sequence, the objective is to learn the conditional probability of $y_{i}$ given previous outputs $y_{<i}$ and input $X$:
\begin{equation}
    P(Y|X) = \prod_{i=1}^N P(y_{i} | X, y_{<i})
    \label{eq:e2e_proba}
\end{equation}

During the inference process, these probabilities are computed from the input test audio frames and the previously predicted phones (not having access to the reference output phones). A beam search algorithm is then applied, with a beam size of 5 and maximum hypothesis length of 30 phone labels for the Lalilo word set, and 130 for the Lalilo sentence set and the adult set. 
No language model is used, as we found that it tends to reconstruct existing words and cover up children's reading mistakes, while we aim at detecting as accurately as possible the phones uttered by children, may they form existing words or not.

Among the existing end-to-end systems, two paradigms stand out in the literature: the CTC-based and the seq2seq architectures. Different methods have been implemented with the latter, using recurrent neural networks and/or attention mechanisms.
Combinations of the two paradigms have also shown great performance. The following sections present the different end-to-end architectures we explored and applied to child speech recognition.

\subsubsection{RNN-CTC model}
\label{sec:ctc}

The CTC paradigm, introduced by~\citet{Graves2006-CTC}, discards the obligation of having an HMM by learning automatically alignments between the input and output sequences.
In phone recognition applications, it aligns the speech frames $X = (x_{1},...,x_{T})$ and their phone sequence $Y = (y_{1},..,y_{N})$, with the condition that $N \leq T$, where $T$ and $N$ are respectively the number of frames in the speech utterance and the length of the phone sequence. 
The length difference between $T$ and $N$ is coped with the addition of a "blank" label, represented by "-", which activation corresponds to the probability of observing no label. During alignment, the system learns a set of possible paths that are phone-label sequences of length $T$ and which probability is the product of all probabilities $p_{k}^{t}$ of observing a label $k$ at time $t$. 
Considering a phone-label sequence $Y$, it can be obtained through a set $\theta_{Y}$ of paths $\pi$, the total probability of Y being defined as follows:
\begin{equation}
    P(Y | X) = \sum_{\pi \in \theta_{Y}} \prod_{t=1}^{T} p^{t}_{\pi_{t}}
    \label{eq:ctc_proba}
\end{equation}

CTC-based networks are trained with gradient descent and the maximum likelihood function, which maximises the log probabilities of the target labels. 
The probability $P(L|X_{1,...T})$ of each individual labelling $L$ is computed with the CTC forward-backward algorithm, which divides the sum over $T$-long paths $\pi_{1,...,T} \in \theta_{L_{1,...,T}}$ into an iterative sum over t-long paths $\pi_{1,...,t} \in \theta_{L_{1,...,t}}$. 
The iterations are then dynamically computed with forward and backward propagation, which makes possible the computation of the many possible paths.

\begin{figure}[!htb]
	\centering
	\includegraphics[scale=.55]{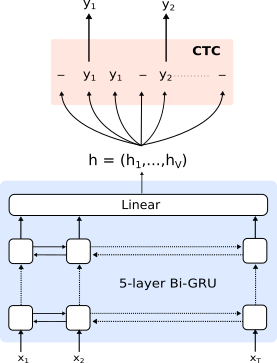}
	\caption{Architecture of the RNN-CTC model}
	\label{fig:RNN_CTC_graph}
\end{figure}

We use, in this work, an end-to-end CTC architecture, named RNN-CTC and shown in Figure~\ref{fig:RNN_CTC_graph}, which is composed of a simple encoder with recurrent neural networks. 
The input features are 40-dimensional MFCC with CMVN, as do~\citet{Bayerl2019-ACH}. The RNNs are composed of Bidirectional Gated Recurrent Unit layers (BiGRU) \citep{Chung2015-GFR}. 
They contain one Bi-GRU input layer with 120 cells, then four Bi-GRU hidden layers with $2 \times 160$ cells in each layer, and finally a linear output layer of dimension 34, corresponding to 33 French phones and the CTC blank label. 
For an increased speed and better concentration of information, we reduced the time resolution of the input by concatenating pairs of consecutive input frames. 
On the output of each hidden layer is applied a 10\% dropout rate. The output layer uses the log-softmax activation function. Models were trained on up to 100 epochs, with an early stopping mechanism linked to the validation loss. Training was done with the Adam optimizer. 
A grid search on the hyper-parameters was done. The adult RNN-CTC models were trained with batch size of 100 and a 9e-5 learning rate. TL models were trained with a batch size of 50 and a 1e-4 learning rate. We used a learning rate scheduler that divided by 10 the learning rate after 2 epochs with no improvement in the validation loss. 
The adult and TL models took an average of 24.3 and 4 hours to train on a single GTX 2080 Ti GPU. They each contain a total of 2.1M trainable parameters.

\subsubsection{Seq2seq LAS model}
\label{sec:las}

At first presented for machine translation tasks, where an input sequence of words in a language needs to be translated to an output sequence of words in another language, seq2seq architectures aim at addressing the problem of performing sequence labelling with variable-length input and output sequences without relying on HMMs like in hybrid DNN-HMM approaches~\citep{Sutskever2014-STS}. 
This framework has shown to be generalisable to many applications: image captioning \citep{Vinyals2015-SAT,Xu2015-SAT}, conversational modelling \citep{Vinyals2015-ANC}, and, of course, acoustic modelling for speech recognition.
ASR tasks are characterised by long input sequences (speech frames) in comparison with other tasks. Unlike CTC-based models, seq2seq models do not assume that the label outputs are conditionally independent of each other, which enables greater freedom but can also cause more frequent confusions. 

Seq2seq models are usually composed of an encoder that extracts information from the audio input and maps it into a variable-length vector, and a decoder that takes this vector and generates an output sequence of tokens (characters, phones or any other speech units), one at a time. 
The use of an attention mechanism~\citep{Bahdanau2015-NMT}, which searches through the encoder output to locate parts of the sequence that are valuable for predicting target tokens, greatly improves the performance. This mechanism generates at each decoder step an attention vector, which conveys some context information from the encoder to the decoder. 

\begin{figure}[!htb]
	\centering
	\includegraphics[width=\linewidth]{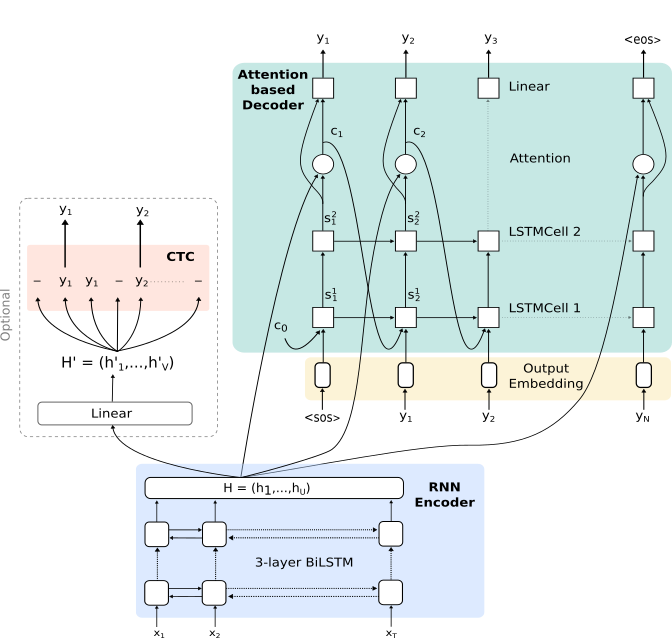}
	\caption{Architecture of the LAS and LAS+CTC models}
	\label{fig:LAS_graph}
\end{figure}

One of the seq2seq models used in this paper is similar to a LAS architecture, recently introduced by~\citet{Chan2015-LAS}. 
Although our architecture, shown on Figure~\ref{fig:LAS_graph}, slightly differs from the initial one, we will in the next sections borrow the name "LAS" for our systems.

The input features to the encoder are 40-dimensional filterbanks with first and second derivatives. We use identical inputs for LAS and LAS+CTC models, following \citet{Watanabe2017-HCA}. 
The encoder contains three layers of bidirectional Long-Short Term Memory (bLSTM) layers, which output a matrix $H = (h_{1}, ..., h_{U})$, where each $h$ has a dimension 256 and $U$ corresponds to the number of audio frames.
A dropout rate of 0.2 is applied to the encoder layers. The decoder takes as input, at each step $i$, the 512-dimension embeddings of the label output $y_{i}$, and passes them through a LSTM layer that outputs a state $s^{(1)}_{i}$ following Equation~\ref{eq:blstm_dec}. 
A second LSTM layer outputs a state $s^{(2)}_{i}$ from the output of the previous layer $s^{(1)}_{i}$ and the previous state of the current layer $s^{(2)}_{i-1}$ (see Equation~\ref{eq:blstm_dec}).
\begin{align}
    \begin{split}
        s^{(1)}_{i} &= \text{bLSTM}(s^{(1)}_{i-1}, y_{i-1}, c_{i-1}) \\
        s^{(2)}_{i} &= \text{bLSTM}(s^{(1)}_{i}, s^{(2)}_{i-1})
    \end{split}
    \label{eq:blstm_dec}
\end{align}
The $c_{i}$ component is the context vector of dimension 512, computed by the attention mechanism from the current state $s^{(2)}_{i}$ and the encoder output $H$ with Equation~\ref{eq:attention}. We use here a simple dot-product attention mechanism, which considers the decoder state $s^{(2)}_{i}$ as the query, and the encoder output $H$ as both the keys and values.
\begin{align}
    \begin{split}
        c_{i} &= \text{DotProductAttention}(s^{(2)}_{i}, H) \\
        &= \text{Softmax}(s^{(2)}_{i}H^{T})H
    \end{split}
    \label{eq:attention}
\end{align}

Finally, a Multi-Layer Perceptron (MLP), composed of two linear layers (size 512) separated by a tanh activation layer, provides the probability $p(y_{i} | X, y_{<i})$ of equation~\ref{eq:e2e_proba}, taking as input the decoder state $s^{(2)}_{i}$ and the attention context $c_{i}$:
\begin{equation}
    p(y_{i} | X, y_{<i}) = \text{MLP}(s_{i}, c_{i})
    \label{eq:mlp}
\end{equation}

Our LAS models are trained with the Adam optimizer, using $l_{2}$ regularisation at a rate of 1e-5, and a learning rate of 1e-4, the latter being halved after each epoch the validation loss did not improve. Scheduled sampling \citep{Bengio2015-SSS} is used to train the LAS models. This method, which consists of feeding the decoder either the reference label or the lastly predicted label, aims at helping the model converge faster and providing more robustness to inference mistakes. The system chooses to input the lastly predicted label instead of the reference label with a fixed probability of 10\%.
All models are trained on 50 epochs, where only the best-validated model is kept. The adult and TL models take an average of 50.7 and 3.9 hours to train on a single GTX 2080 Ti GPU. The final models each contain a total of 7.6M parameters. 




\subsubsection{Seq2seq LAS+CTC model}
\label{sec:las-ctc}

The simple attention mechanism used in the LAS is too flexible for ASR tasks, because it allows non-sequential alignments, while speech recognition inputs and outputs are generally sequential~\citep{Watanabe2017-HCA}. 
Another characteristic of speech recognition is that the input and output sequences can vary greatly in length, especially with reading learners with different reading levels, which makes the alignments more difficult to track.
Since CTC enables to compute constrained monotonic alignments, it was proposed by~\citet{Watanabe2017-HCA} to combine both paradigms into a hybrid CTC/attention-based system. During training, a CE+CTC multi-objective learning method combines the two losses. Two outputs, one from the encoder CTC and the other from the attention-based decoder, can generate output sequences.

The CTC-attention hybrid models follow the same architecture and use the same hyper-parameters as the LAS models described in the previous section. It also uses the same input features, filterbanks of dimension 40, with first and second derivatives, as do~\citet{Watanabe2017-HCA}. Scheduled sampling is applied with a 10\% rate.
The CTC loss is combined with the attention loss following Equation~\ref{eq:loss}, with $\lambda = 0.2$. Other values of $\lambda$ (0.5, 0.8) gave worse results, as found by~\citet{Watanabe2017-HCA}.
\begin{equation}
    \text{loss} = \lambda \times \text{loss}_{CTC} + (1 - \lambda) \times \text{loss}_{CE}
    \label{eq:loss}
\end{equation}


The attention-based decoder and encoder CTC outputs are taken separately to generate two sets of hypothesised label sequences.  
Two scores will thus be presented for the LAS+CTC model, which will be denoted as \textit{LAS+CTC enc} and \textit{LAS+CTC dec} for the encoder CTC and attention-based decoder outputs, respectively.

\subsubsection{Seq2seq Transformer model}

Presented by~\citet{Vaswani2017-AIA} and adapted to speech recognition by~\citet{Dong2018-STR}, the Transformer model follows a Seq2seq encoder-decoder architecture, but relies solely on attention mechanisms, instead of recurrent neural networks in classical Seq2seq systems. 
The recurrence, essential to extract position information in speech frames, is replaced by positional encodings concatenated to the input encodings, as well as multi-head self-attention mechanisms and position-wise feed-forward neural networks in the encoder and decoder blocks. 
Discarding the need of recurrent neural networks enables to compute dependencies between each pair of positions at once, instead of one by one. It allows for faster training and more parallelisation in comparison with LAS systems presented in previous sections.
%
%
\begin{figure}[!htb]
	\centering
	\includegraphics[width=\linewidth]{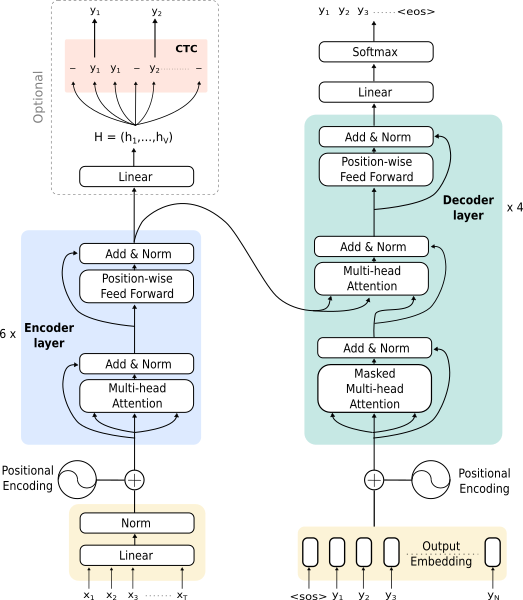}
	\caption{Architecture of the Transformer and Transformer+CTC models}
	\label{fig:Transf_graph}
\end{figure}


Figure~\ref{fig:Transf_graph} presents the architecture of our Transformer, with the optional CTC module (see section~\ref{sec:transf-ctc}).

The Transformer models' encoder takes as input the audio sequence $X = (x_{1},...,x_{T})$, where $x_{i}$ is a 80-dimensional filterbank vector. The inputs are processed with linear and normalisation layers of size $d_{\text{model}} = 256$. Positional encodings give information about the relative position of the tokens in the sequences, replacing the recurrence or convolution used in classical models. These encodings have the same dimension (256) as input/output embeddings to ease the summation between the two vectors. 
As in literature \citep{Vaswani2017-AIA, Dong2018-STR, Karita2019-ACS}, we use sinusoidal positional encodings, defined as: 
\begin{align}
    \begin{split}
        \text{PosEnc}_{(pos,2i)} &= \sin(pos/10000^{2i / d_{\text{model}}}) \\
        \text{PosEnc}_{(pos,2i+1)} &= \cos(pos/10000^{2i / d_{\text{model}}})
    \end{split}
    \label{eq:pos_enc}
\end{align}
where $pos$ is the audio frame or label position, and $i$ the $i$-th dimension of the positional encodings, with $0 \leq i < d_{\text{model}}$. 

The encoder contains six layers, each layer being composed of two sub-layers: a multi-head self-attention module, and a position-wise feed-forward fully connected neural network module. Each of these modules is followed by a normalisation layer with residual connection~\citep{He2016-DRL}. 
The first sub-layer contains a multi-head scaled dot-product attention~\citep{Vaswani2017-AIA}, which relates different positions of inputs to create more valuable representations. Each of the $i \in \{1..h, h=4\}$ heads follows Eq.~\ref{eq:attention_transf}, where $Q_{i}$, $K_{i}$ and $V_{i}$ represent respectively the $i$-th linear projection of the query, keys and values, and $d_{k} = 256$ is the dimension of $Q$, $K$ and $V$. The scaling by $1/\sqrt{d_{k}}$ restrains the dot product from growing too large when $d_k$ is large, which would give very small gradients with the softmax function. The final output of this module is a linear projection of the concatenation of these 4 heads.
The second sub-layer is composed of a position-wise feed-forward neural network that consists in two linear layers separated by a ReLU activation, and is applied to each position separately. 
\begin{align}
    \begin{split}
        \text{head}_{i} &= \text{ScaledDotProductAttention}(Q_{i}, K_{i}, V_{i}) \\ &= \text{Softmax}\left(\frac{Q_{i} K_{i}^{T}}{\sqrt{d_{k}}}\right)V_{i}
    \end{split}
    \label{eq:attention_transf}
\end{align}
%


The inputs of the decoder, which are the reference phone-labels during training, and the previously decoded phone-labels during inference, are offset by one position with the start-of-sequence (<sos>) token, and go through an embedding layer of dimension 256. As for the encoder inputs, positional encodings are computed and summed to the decoder's inputs. Unlike LAS models, which decode token by token, the Transformer's decoder takes as input the whole utterance at the same time, enabling to parallelise and decode faster. This characteristic however prevents from using scheduled sampling as defined by \citet{Bengio2015-SSS}, where the previously predicted token is needed to predict the current token.

The four layers of the decoder each contains three sub-layers. The first is the same self-attention module as in the encoder, with the difference that it is masked. 
This mask, combined with the offset by one position of the phone-label embeddings serving as inputs to the decoder, ensures that for a position $i$, the predictions depend only on prior $i-1$ positions, sole known positions in inference mode.
The second sub-layer is another multi-head attention module, which functions as the attention mechanisms in other seq2seq systems: the encoder stack outputs the keys and values, while the query is provided by the prior sub-layer of the decoder.
The third sub-layer is a position-wise feed forward neural network, which is identical to the second sub-layer of the encoder.

The decoder output are finally processed through a classification network composed of a linear projection and a softmax operation, which yield the final phone-label sequence.

The Transformer models are trained with the Adam optimizer with $\beta_{1}=0.9$, $\beta_{2}=0.98$ and $\epsilon=$ 1e-9. We use the same specific learning rate scheduler as \citet{Vaswani2017-AIA} for training, defined in Equation~\ref{eq:lrate}, with $\text{warmup}\_\text{steps}=4000$. 
\begin{equation}
    \text{lr} = d_{\text{model}}^{-0.5} \cdot \text{min}(\text{step}^{-0.5}, \text{step}\cdot \text{warmup}\_\text{steps}^{-1.5})
    \label{eq:lrate}
\end{equation}

Models are trained on 100 epochs and we score with the best validated model. The adult and TL models respectively take in average 30 and 3 hours to train on a single GTX 2080 Ti GPU. The final model contains a total of 14.3M parameters.

\subsubsection{Seq2seq Transformer+CTC model}
\label{sec:transf-ctc}

In the same objective as with the LAS+CTC system, we can optionally use the CTC objective function to improve the Transformer model training and inferring processes, as has been shown by~\citet{Karita2019-ITB}. In the same way, the CTC and attention objectives can be combined with Equation~\ref{eq:loss}. 

The Transformer+CTC models use the same inputs, filterbanks of dimension 80, as do~\citet{Karita2019-ITB,Karita2019-ACS}. The models were trained with the same process as the Transformer, and the combination of objectives worked the best with $\lambda = 0.3$ as in~\citet{Karita2019-ITB}. 
The adult model trained during 100 epochs, while 10 epochs were sufficient to train the TL model, resulting in a very short training time for the latter (below one hour).

In the same way as the LAS+CTC, the Transformer+CTC model has two outputs, one from the encoder with the CTC function and one from the decoder with the attention mechanism. These outputs are denoted \textit{Transformer+CTC enc} and \textit{Transformer+CTC dec}, respectively, in the following.


\section{Evaluation}
\label{sec:results}

With the implementation of a phone recognition for a reading assistant application, we do not aim at reconstituting and correcting words based on the detected phones, but at transcribing accurately what the child has read, including potential phone-level reading mistakes. 
Therefore, we do not measure performance with the classical WER but with a Phone Error Rate (PER).
The PER metric considers all mismatches between the recogniser hypothesis and the manual phone-level annotated reference, and is defined in Equation~\eqref{eq:PER}, with $C$, $I$, $S$, $D$ respectively referring to the number of correct detections, insertions, substitutions and deletions.

\begin{equation}
    \centering
    \text{PER} = \frac{I + S + D}{C + S + D}
    \label{eq:PER}
\end{equation}

\subsection{Comparison of architectures on adult speech}

Adult models were trained with about 150 hours (see Table~\ref{tab:info_corpus_CV}). 
We score their performance with the PER metric on the Test CV adult speech set, as displayed in Table~\ref{tab:adult_results}. 

\begin{table}[!htb]
  \caption{PER (\%) obtained with adult-trained acoustic models tested on adult speech}
  \label{tab:adult_results}
  \centering
  \begin{tabular}{l | c }
    \toprule
    Model & Test CV \\
    \midrule 
    TDNNF-HMM \textit{(baseline)} & 23.5  \\
    \midrule
    RNN-CTC & 16.1 \\
    \hdashline[1pt/1pt]
    LAS & 12.6 \\
    LAS+CTC enc & 16.7 \\
    LAS+CTC dec & 11.9 \\
    \hdashline[1pt/1pt]
    Transformer & \hphantom{1}\textbf{7.5}\\
    Transformer+CTC enc & 11.9 \\
    Transformer+CTC dec & \hphantom{1}8.0 \\
    \bottomrule
  \end{tabular}
\end{table}

We observe that the models' performance follows their chronological appearance in the literature: the worse is the baseline, the TDNNF-HMM model with a PER of 23.5\%, and the best is the Transformer with a PER of 7.5\%. 
In addition to being simpler to train, end-to-end models all perform significantly better than the baseline hybrid TDNNF-HMM model (from 29\% to 68\% relative improvement on the PER). 
Attention mechanisms show to be particularly efficient: the LAS, LAS+CTC dec and Transformer+CTC dec systems give lower PER than the RNN-CTC or the TDNNF-HMM, and the Transformer, which is solely based on attention, reaches the lowest PER. 
The CTC objective function brings valuable information for training (LAS+CTC dec is better than LAS) but inferring hypothesis sequences with the CTC output is not as accurate as using the attention-based decoder output (LAS+CTC enc is significantly worse than LAS+CTC dec). 
We did, however, not succeed in exploiting the previously shown potential of the CTC objective function for the Transformer, resulting in a slightly worse performance for model Transformer+CTC dec in comparison with the Transformer.
As seen on LAS+CTC models, using the encoder CTC output of the Transformer+CTC for inferring sequences does not provide comparable performance with using the decoder output.

\subsection{Comparison of models on child speech}

In this section we test the adult-trained model on child speech, and compare their performance with models trained with transfer learning. 
We train TL models from adult models, so they benefit from the 150 hours of adult speech, as well as the 13 hours of child speech. Table~\ref{tab:TL_results} displays the PER scores for all architectures, when trained only on adult speech (first column) or adapted with child speech (second column). 

\begin{table}[htb]
  \caption{PER (\%) obtained with acoustic models without (adult-trained) and with TL, tested on child speech}
  \label{tab:TL_results}
  \centering
  \begin{tabular}{l | c c}
    \toprule
    Model & Without TL & With TL \\
    \midrule
    TDNNF-HMM \textit{(baseline)}& \textbf{45.9} & 30.1 \\
    \midrule
    RNN-CTC & 59.4 & 32.4 \\
    \hdashline[1pt/1pt]
    LAS & 77.9 & 33.9 \\
    LAS+CTC enc & 63.5 & 32.7 \\
    LAS+CTC dec & 72.9 & 30.7 \\
    \hdashline[1pt/1pt]
    Transformer & 76.6 & 28.5\\
    Transformer+CTC enc & 60.0 & 29.3 \\
    Transformer+CTC dec & 70.9 & \textbf{28.1} \\
    \bottomrule
  \end{tabular}
\end{table}

The six systems could in theory be also trained directly on the child training set to obtain child-adapted acoustic models. However, any speech recognition system necessitates a proper amount of data to correctly learn speech representations, especially the latest end-to-end architectures, and we have only 13 hours of child training data. 
Child models inevitably gave us incoherent and poor results, which will thus not be presented here.

\subsubsection{Adult models on child speech}

The difference in score for adult models on adult speech (Table~\ref{tab:adult_results}) and child speech (Table~\ref{tab:TL_results}) is drastic: all models loose in average 51 points of PER. 
These extreme gaps in performance may be explained by the acoustic and prosodic difference between adult and child speech. 

Contrary to what has been observed for adult models when tested on adult speech, the TDNNF-HMM performs significantly better than its end-to-end counterparts when tes\-ted on child speech: its PER score is doubled, while end-to-end systems' scores suffer from multiplicative factors that go from 3.8 (LAS+CTC enc) to 10.2 (Transformer). 
In \cite{Chan2015-LAS}, end-to-end models were shown to perform less well on short utterances, which is in line with our results.
The CTC objective function helps the attention module during training, as show the comparisons between LAS and LAS+CTC dec models, and between Transformer and Transformer+CTC dec models. Additionally, inferring phone sequences with the encoder CTC output shows better performance than with the attention-based decoder output for both LAS and Transformer models. Finally, the RNN-CTC model obtains a significantly better score than attention-based models. These observations suggest that attention-based systems necessitate matched training and testing sets, while CTC-based systems are able to cope with slight mismatches.

\subsubsection{TL models on child speech}

At first sight, we observe that TL models perform drastically better than adult-trained models, showing the positive effect of transfer learning even with a quantity of data as small as 13 hours. We observe a mean relative gain of 50.9\% between scores in the first and second column. 
The attention-based systems (LAS and Transformers) benefit the most from the transfer learning, and the most spectacular improvement can be seen on the Transformer's performance, corresponding to a relative reduction of 62.8\% of the PER. 
Models that do not rely on attention, i.e. TDNNF-HMM and RNN-CTC, benefit from TL to a lesser extent (\textit{resp.} 34.4\% and 45.5\% relative improvements). The TL models obtain PER scores that are included in a relatively small range (28\% - 34\%), which is interesting considering the very wide range of results for models without TL (45\% - 78\%). 
This suggests that independently of the architecture, and due to the small amount of target data, the efficiency of transfer learning might reach a limit when confronted to arduous tasks like speech recognition for children learning to read.

The attention-based systems, whose attention mechani\-sms trained on adult sentences were not able to adapt to the diverse utterance lengths and mismatched prosodic characteristics of child speech, show to reach, thanks to the transfer learning, comparable performance with the non-attention-based systems. 
The child speech training data, containing equal proportions of isolated words and sentences, empowers the attention mechanisms to adjust to its specificities. In this way, the models based on the Transformer architecture outperform the TDNNF-HMM baseline with relative improvements up to 6.6\%. 
The best architecture is the Transformer+CTC with the decoder output, reaching a 28.1\% PER. The systems that use RNNs are less efficient than TDNNF-HMM and Transformers, suggesting that they need more data to gain precision on extracting relevant information in child speech.
These results concur with the global ranking between hybrid DNN-HMM, RNN-based seq2seq and Transformer architectures presented in~\citet{Karita2019-ACS}.

If we focus on the influence of the CTC, we observe the same patterns for LAS and Transformers architectures, although to various extents. It shows highly useful for training the LAS+CTC, improving by a relative 9.4\% the score of the LAS+CTC dec model in comparison with the LAS model, and brings a lower improvement between Transformer and Transformer+CTC dec models (only 1.4\% relative). 

\section{Discussion}
\label{sec:discussion}
We presented different architectures for acoustic modelling and compared their performance when applied to the specific case of child speech with limited data. 
Although PER scores obtained in the previous section may still seem high, it is important to note that the child speech recordings can contain high levels of babble noise, typical in children environments. The mean SNRs of the Lalilo training, validation and testing sets are 20.9 dB, 20.1 dB and 20.6 dB, respectively (see Table~\ref{tab:info_corpus_Lalilo}). These values are much lower than those of the Common Voice dataset (\textasciitilde34.3 dB, see Table~\ref{tab:info_corpus_CV}), which has been gathered in relatively calm environments. While the best model obtains a 28.1\% PER on all recordings, it performs significantly better on clean recordings (i.e. SNR~$\geq$~20~dB), reaching a 20.0\% PER. 
The presence of babble noise on recordings with SNR~$\leq$~10~dB degrades the PER by 124.0\% relative in comparison with clean ones. The robustness of the system to babble noise could be improved with data augmentation and multi-condition training techniques~\citep{Gibson2018-MCD,Airaksinen2019-DAS}: it would most certainly improve the global performance by providing additional child speech data, as well as better training the model to recognise speech in noise.

In the scope of phone recognition on young readers' speech, we will in this section further explore the use of these models for evaluating specific reading tasks, i.e. isolated words and sentences. 
An analysis of the behaviour of the Transformer+CTC dec model in the presence of reading mistakes will also be carried, with a focus on the beneficial influence of the CE+CTC multi-objective training for detecting common sentence-level reading mistakes.

\subsection{Application to specific reading tasks}

\begin{figure}[!htb]
	\centering
	\includegraphics[scale=.32]{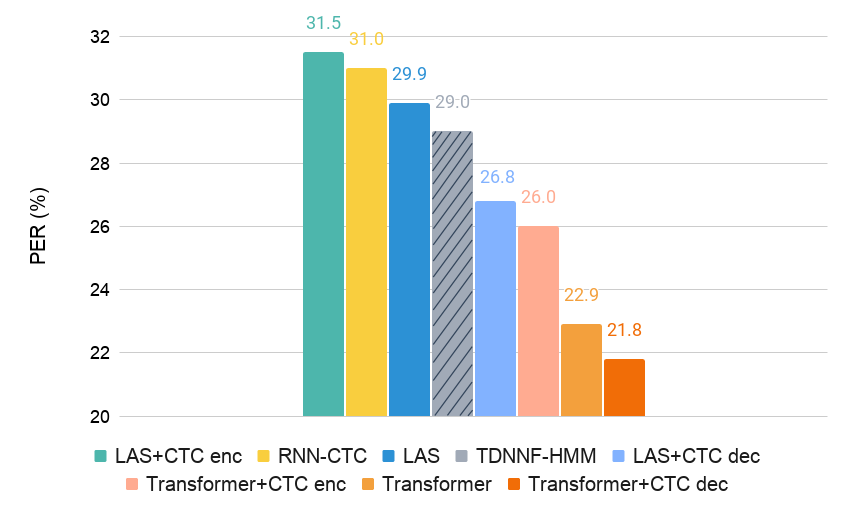}
	\caption{PER (\%) of all TL models on Test S (sentences)}
	\label{fig:WS_study_S}
\end{figure}
\begin{figure}[!htb]
	\centering
	\includegraphics[scale=.32]{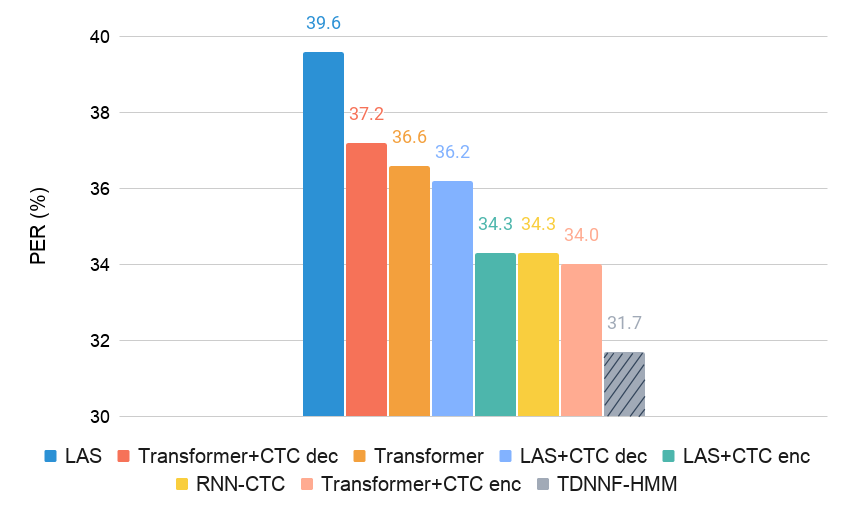}
	\caption{PER (\%) of all TL models on Test W (isolated words)}
	\label{fig:WS_study_W}
\end{figure}

Children learning to read are typically offered different reading tasks, based on their reading level or on the skill they have to improve.
We focus here on two reading  tasks: isolated words and sentences. 
Knowing that acoustic model architectures might behave differently depending on the length of the utterance or the number of words, we take an interest in detailing each architecture's behaviour on these two tasks. 
Figures~\ref{fig:WS_study_S} and~\ref{fig:WS_study_W} display the models' performance on the child speech test sets, respectively Test S (sentences) and Test W (isolated words).


On sentences, Transformers obtain far better scores than the others, outperforming the TDNNF-HMM baseline by relative improvements from 10.3\% to 24.8\%. The Transformer+CTC dec obtains the lowest score of 21.8\%.
A clear tendency can be observed on Fig.~\ref{fig:WS_study_S} for models using multi-objective CE+CTC training. It shows to bring improvement: the LAS+CTC dec model shows a relative gain of 10.4\% over the LAS model, the Transfor\-mer+CTC dec of 4.8\% over the Transformer model. For both architectures, the CTC encoder output seems to bring significantly worse results than the decoder output that relies on attention. This observation, combined with the poor score of the RNN-CTC on sentences, shows that the CTC objective on its own has difficulties for recognising sentences, probably due to a high complexity in the alignment of sentences uttered by reading learners, i.e. including reading miscues such as repetitions. 

The scores on isolated words (Test W) are significantly higher than on sentences, suggesting that the task is quite arduous. 
This difficulty is most probably brought by the short length of utterance, as well as the higher proportion of reading mistakes (in Test W, 51.5\% of phones belong to an utterance that contains at least one reading mistake, versus 36.0\% in Test S, see Table~\ref{tab:info_corpus_Lalilo}). 
On isolated words, the baseline hybrid TDNNF-HMM architecture is far from being outranked, reaching a PER of 31.7\%, which lies 2.3\% absolute below the second-best model. 
Very clear patterns can be observed on Figure~\ref{fig:WS_study_W}, with well-defined groups of PER. The three models that use the CTC as output (LAS+CTC enc, RNN-CTC and Transformer+CTC enc) form a first group with PERs between 34.0\% and 34.3\%, which correspond to 7.3\% to 8.2\% relative degradation in comparison to the baseline. We can infer from this that the CTC takes over the attention mechanisms when the utterances are too short, and manages to align the phones properly with a very small context. 
We then see a second group, which includes the two models that are trained with the CTC objective and use the decoder output (Transformer+CTC dec and LAS+CTC dec), as well as the Transformer model. This floor corresponds to 14.2\% to 17.4\% relative augmentation over the baseline. Contrary to sentences, the CE+CTC multi-objective training has mitigated effect on LAS and Transformer models. 
While it is still highly efficient for LAS models (8.6\% relative improvements between LAS and LAS+CTC dec models), it slightly degrades the performance of the Transformer (-1.6\% relative). 
The worse result is obtained by the LAS model, with a 24.9\% relative augmentation of the PER in comparison with the baseline. These four models ranking last on isolated words are all using attention-based decoder to infer phone sequences. 
It shows that attention mechanisms have difficulties inferring labels from very short audio sequences. This result echoes the results of~\citet{Chan2015-LAS}, where their error rate significantly rises for 1-word utterances.

\subsection{Application to non-proficient readers' speech}

Aiming at phone recognition for young readers, our system will inevitably encounter many reading mistakes, and must correctly recognise the uttered phones, independently from the number of reading mistakes and the child's reading level. However, the presence of reading mistakes can severely hurt the model's accuracy. In this section, we evaluate the Transformer+CTC dec performance when confronted to reading mistakes, and analyse further the effect of the CTC objective function towards repetitions, very common mistakes for young readers.

\begin{figure}[!htb]
	\centering
	\includegraphics[scale=.35]{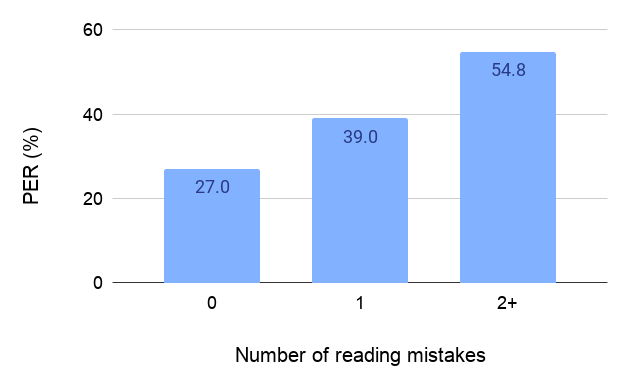}
	\caption{PER (\%) of the Transformer TL model on Test W as a function of the number of reading mistakes in the utterance}
	\label{fig:Err_study_w}
\end{figure}

\begin{figure}[!htb]
	\centering
	\includegraphics[scale=.28]{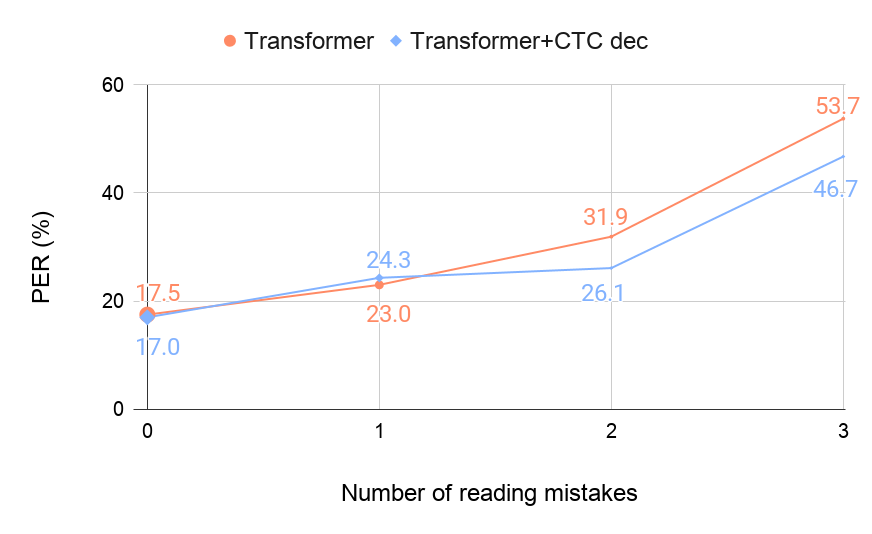}
	\caption{PER (\%) of the Transformer TL model on Test S as a function of the number of reading mistakes in the utterance}
	\label{fig:Err_study_s}
\end{figure}

Figures~\ref{fig:Err_study_w} and~\ref{fig:Err_study_s} display the evolution of the PER scores obtained by the Transformer+CTC dec model in function of the number of reading mistakes contained in the utterances, for Test W and Test S, respectively. 
Figure~\ref{fig:Err_study_s} additionally includes the results of the Transformer model, which will later serve as comparison to study the influence of the CTC function.

In Test W, we count the number of mistakes at the word-level, in terms of phones: for example, a word with a substituted phone is counted as one mistake, a word with a hesitation and an inserted phone belongs in the 2-and-more mistakes category, and a false start or a repeated word where the child pronounces more than two phones is also classified as 2+ mistakes. 
As indicated in Table~\ref{tab:info_corpus_Lalilo}, 48.5\% of phones uttered in Test W belong in the no-mistake category, while 1 mistake and 2-and-more mistake categories include respectively 26.0\% and 25.5\% of the phones.  
In Test S, on the other hand, we count the number of mistakes at the sentence-level, in terms of words: each misread word counts as a mistake, a false start, a repeated word or a word containing a hesitation as well. 
Test S contains 64\% of phones belonging to correctly read utterances (see Table~\ref{tab:info_corpus_Lalilo}), and other categories contain (in ascending order) 20\%, 6.8\%, and 9.2\% of the phones uttered in Test S. On Figure~\ref{fig:Err_study_s}, the size of the points is proportional to the number of reference phones in each category.

\begin{figure*}
    \centering
	\includegraphics[width=0.9\linewidth]{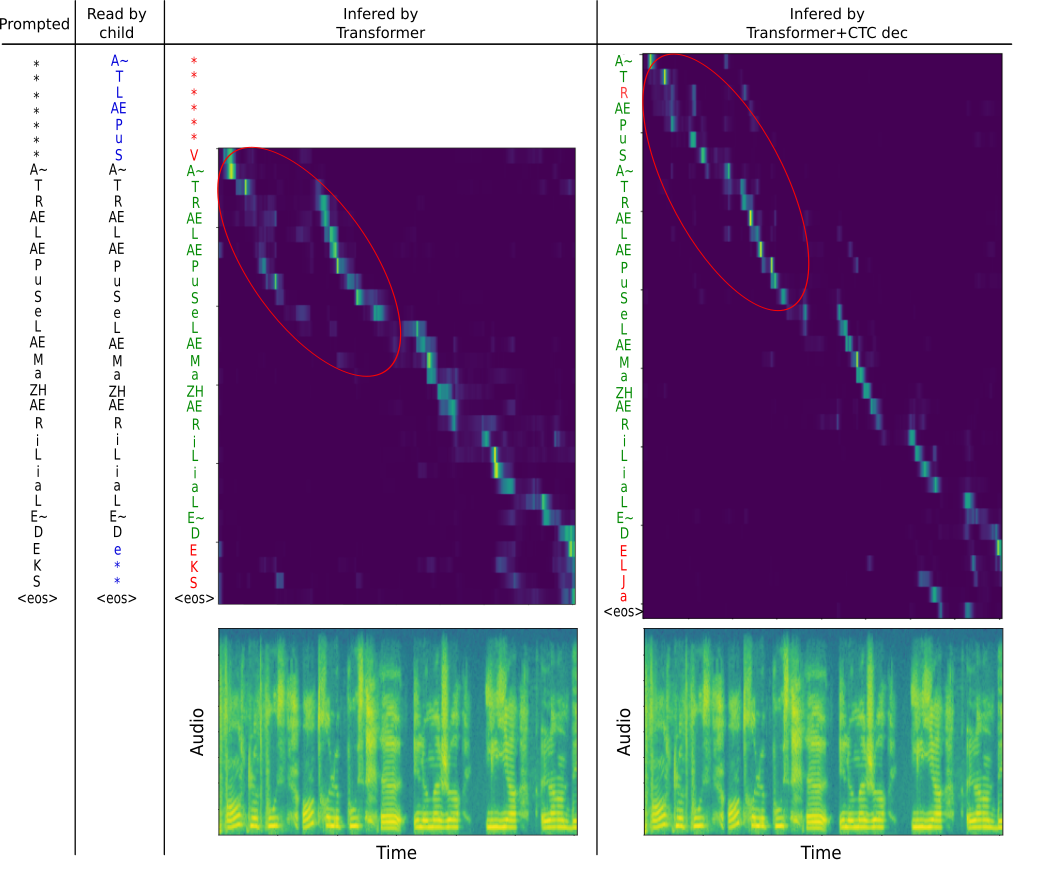}
    \caption{Phone-level prompted text, actual text read by the child and transcriptions obtained with models Transformer and Transformer+CTC dec of an example utterance. Child's reading mistakes (repetitions, phone substitutions and phone deletions) are in blue, and correct detections and errors in models' transcriptions in green and red, respectively. Attention weights captured between the encoder and decoder of Transformer and Transformer+CTC dec models are displayed, as well as the spectrograms for time reference.}
    \label{fig:Att_repet}
\end{figure*}

We can easily notice the tendency following which the more reading mistakes are present in the utterance, the more degraded is the PER, for both isolated words and sentences. 
On Test W, the presence of a single reading mistake augments the PER by 44.4\% relative in comparison with correct readings. More mistakes occurring in an utterance degrade the PER in the same significant extent: a relative degradation of 40.5\% is observed between 1 and 2+ mistakes. 
On Test S, the evolution is also striking: for both Transformer and Transformer+CTC dec models, the PER drastically increases when the utterances contain reading mistakes. 
Analysing further the Transformer+CTC dec performance, we see that going from correct reading to reading with one mistake brings a 42.9\% relative degradation. The score for utterances with two mistakes is only slightly higher than with a single mistake, while it highly augments for utterances with more than three reading mistakes (78.9\% relative augmentation over the category 2).

This striking phenomenon can be explained by the characteristics of children's typical reading mistakes (detailed in section~\ref{sec:typical_mistakes}). 
Beginner readers tend to read with a slow speech rate and prosody: hesitations inter- or intra-word and phones of longer duration, for instance. 
These peculiarities can cause the attention mechanisms to lose focus and miss out on some phones. 
Additionally, substitutions of phones can sometimes give unusual combinations of phones that do not exist or are rarely present in the language, which the model is not trained to recognise. 
Finally, repetitions are usual reading mistakes (in sentences in particular), as the students often start back from the beginning of the sentence when they have difficulties, and can cause great confusion.
The attention mechanisms risk in this case to attend the wrong occurrence of the word, or to consider two occurrences as a single one. This is problematic when the child reads a word first time with a mistake, then a second time self-correcting, which is common, because the attention could confuse the two occurrences and miss the mistake.

Fig.~\ref{fig:Att_repet} displays, for an utterance that well represents the type of reading mistakes a young reader can do, the phone-level prompted text and actual text read by the child, as well as the transcriptions obtained with Transformer and Transformer+CTC dec models. 
Additionally, it presents the weights extracted from the attention module that links the encoder and decoder of the two systems. The spectrogram of the utterance is also shown for time reference. 
By comparing the prompted text and the text read by the child, we can see that the child repeats the first three words (the repeated phones are in blue on Fig.~\ref{fig:Att_repet}), and makes a mistake when reading the first word (\textit{A$\sim$ T} instead of \textit{A$\sim$ T R AE}).
A first observation is that the Transformer does not detect the first attempt of the child, while the Transformer+CTC dec does, although with a substitution of a phone ($R$ instead of $L$, in red on Fig.~\ref{fig:Att_repet}). 
The parallel diagonals that are pointed out by a red ellipse on the Transformer attention weights show that the model confuses the two occurrences of the repeated words into one.
On the contrary, the Transformer+CTC dec attention weights form a single diagonal, correctly detecting the two occurrences as separate. This success may be explained by the influence of the CTC objective function, which constrains the attention to be monotonic, enabling the model to correctly detect repetitions.

Another observation is that the child makes several word-level reading mistakes in the last word of the sentence: a phone substitution and two phone deletions (\textit{E$\sim$ D e} instead of \textit{E$\sim$ D E K S}). 
The word read by the child does not exist in the French language and is not represented in the training data. The Transformer unexpectedly detects the correct reading, which means it misses the child's mistake. 
This is due to the presence of this same content in the child speech training data, but read correctly by other students, which has caused the Transformer to predict the learnt existing word. 
On the other hand, the Transformer+CTC dec model does detect a reading mistake in the last word, forming a non-existing word, although not the one uttered by the child. We can infer that since it is able to form non-existing words, it might show more robustness to children reading mistakes, as shown in Fig.~\ref{fig:Err_study_s}.

The issues that attention mechanisms face when confronted to repetition or non-existing words can cause problematic errors when transcribing young readers' mistakes. The child training data, containing only correct readings, prevents the models to be trained to handle readings with mistakes. 
A perspective for improving our model's performance in the presence of reading mistakes is to add some representations of these mistakes in the child training data. 
It would enable the model to better detect incongruous substitutions and insertions of phones, as well as repetitions, false starts and hesitations. However, this method would require manual annotation at the phone-level of a significant quantity of data, which is an arduous and costly task. 
Synthetic errors could partly remedy the issue: although word-level mistakes might be thorny to create, utterances with sentence-level mistakes such as repetitions, false starts and hesitations could be automatically created and added to the training data. Finally, our Transformer models could gain robustness on child speech peculiarities during inference by using scheduled sampling methods, adapted to the specific Transformer structure, during training \citep{Mihaylova2019-SST,Zhou2019-IGT}.

\section{Conclusion}

Speech technologies are nowadays widening their usage domains, including numerous applications designed for children. However, the speech recognition systems' performance is significantly lower on child speech than on adult speech. In particular, educational numerical resources for children learning to read, could immensely benefit from accurate speech recognition to detect reading mistakes. 

End-to-end architectures have proved their ability to outperform hybrid DNN-HMM approaches for ASR. 
In this work, we apply end-to-end architectures to child speech with a limited amount of child data, and show that, with the help of transfer learning strategies, a Transformer+CTC can reach a 28.1\% PER, and outmatch a TDNNF-HMM model by 6.6\% relative, as well as other end-to-end architectures (RNN, LAS) by 8.5\% to 17.1\% relative. 
We specifically study CE+CTC multi-objective training on diverse end-to-end architectures, which shows to bring significant improvements. 
We find a degradation of performance for attention-based models when the utterance length is reduced, while TDNNF-HMM and CTC-based models seem to better handle very short utterances. 
Finally, we compare the performance of Transformer and Transformer+CTC models in presence of reading mistakes, and show that CE+CTC multi-objective training indeed constrains the attention to be more monotonic, which enables the Transformer+CTC model to better detect common young readers' mistakes.

Our detailed studies provide valuable insights on the end-to-end architectures remaining weaknesses for transcribing young readers' speech. 
Using different speech units, such as syllables or characters, on their own or in combination with phones, could bring complementary information on reading learners' speech. 
Having seen that the small amount of child data could limit the potential of transfer learning, diverse data augmentation methods could be investigated. 
The problem of noise could in particular be tackled with data augmented with child-adapted babble noise. Creating synthetic child training data with common reading mistakes, such as repetitions, false starts, or hesitations would enable the models to better handle children mistakes. 


\bibliographystyle{cas-model2-names}

\bibliography{mybib}

\end{document}